\documentstyle[12pt]{article}
\def \eps{\varepsilon}
\def \ln{{\rm ln}}
\def \d{\partial}
\author{Ya. Urzhumov}
\title{Special relativistic spherically symmetric Lagrangean hydrodynamics
 from General Relativity}

\begin{document}
\begin{center}
\Large Special relativistic spherically symmetric Lagrangean hydrodynamics\\
 from General Relativity \\
\normalsize
\vskip 1cm
{\bf Yaroslav Urzhumov \footnote{e-mail: Yaroslav.Urzhumov@itep.ru } } \\
\vskip 1cm
Institute for Theoretical and Experimental Physics \\
B. Cheremushkinskaya 25, Moscow, 117259 Russia
\end{center}
\vskip 2cm
\centerline{\bf Abstract}
\vskip 1cm
We re-derive hydrodynamical equations in General Relativity (GR) in the comoving reference frame for spherical symmetry
 and obtain from
them
the well-known but not explicitly derived  Lagrangean equations in Special Relativity (SR), that
is, for the case of weak gravitational fields. Explicit formulae are presented which
relate
General Relativistic independent variables, Lagrangean mass coordinate and
variables in the lab-frame. Conversion of one set of variables into another
requires knowing the solution to the set of equations. These formulae allow
one to translate the solution to the exact set in General Relativity into the
form in which Special Relativistic solutions are usually obtained. This is
applicable for comparison of SR-numerical simulations of collapses, GRBs and
Supernovae explosions with more precise GR-simulations.

\newpage

 \section{General relativistic hydrodynamics in the comoving reference frame}

\par\indent

{\it
A detailed derivation of the hydrodynamical equations for the
spherically symmetric motion in 3+1 dimensions can be found in the paper
[5] (Liebend\"orfer, Mezzacappa \& Thielemann), along with  relativistic Boltzmann equation and its forms.
However, the
relation of General Relativistic hydrodynamical equations with Special Relativistic ones, such as used
without derivation by Daigne \& Mochkovitch [2], has remained unclear. We
fill this blank with a comprehensive derivation of the basic equations in [2]
starting from the very Einstein equations.}

 We will work in the reference frame moving with (baryonic) matter and suppose that the metric tensor has been diagonalised,
adopting the
following (already spherically symmetric) form of the interval:
$$ ds^2= e^{2\varphi} dt^2 -e^{2\lambda} dR^2 -r^2 d\Omega^2, \eqno(1)$$
where $d\Omega^2 \equiv d\theta^2+ \sin^2\theta d\phi^2 $ is a solid angle
differential.
\def \phi{\varphi}
 The quantities $\phi,\lambda, r $ are the
functions of independent variables $R,t$, which are correspondingly the arbitrarily chosen radial
and universal time coordinates. We have already  set $c$ to be 1.

Jacobian of these 4-coordinates is $\sqrt{-g}=e^{\phi+\lambda}r^2
|\sin\theta|$, and jacobian of spatial coordinates equals
$\sqrt{\gamma}=e^\lambda r^2|\sin\theta|$, hence,
a 3-dimensional infinitesimal volume  element integrated over $\phi,
\theta$ equals
$$ dV=e^\lambda 4\pi r^2 dR, \eqno(2) $$
and $r(R,t)$ has the geometrical sense that the length of circle of radius $r$
is $2\pi r$.

Now we have to put some physical sense to the coordinate $R$ in order to fix
its definition (one can apply a conversion $R\to \tilde R$ staying in the comoving
frame and only changing $\lambda\to\tilde\lambda$, so  $R$ is not fixed). We relate this coordinate with the baryon
density, so that $dR/G \equiv dm_b$ is the baryon mass inside the layer
$dR$:
 $$ \rho_b dV = \rho_b e^\lambda 4\pi r^2 dR = dm_b = {1\over G}dR. \eqno(3) $$
In the following we put $G=1$ (so that mass and length are of the same dimension)
and denote by prime the $\d /\d  R$ derivative.

In our hands are the baryon number conservation law
$$ (\rho_b u^\mu)_{;\mu}=0 $$
and the Einstein equations
$$ R_{\mu\nu}-{1\over 2}Rg_{\mu\nu}=8\pi T_{\mu\nu}.$$
In our comoving reference frame the 4-velocity equals  $u^\mu=(e^{-\phi},
\vec 0)$, and using the formula for $ \Gamma^\mu_{\nu\mu}={\d  \ln
\sqrt{-g} \over \d  x^\nu}$ we get
$$ (\rho_b u^\mu)_{;\mu}=(\rho_b u^\mu)_{,\mu}+\Gamma^\mu_{\nu\mu}\rho_b u^\nu=
e^{-\phi}(\dot \rho
+\rho\dot\lambda+2\rho\frac{\dot r}{r})=0,$$
that is,
$$ {\dot \rho \over \rho}=-(\dot\lambda+2\frac{\dot r}{r}). \eqno(4)$$
(We omit the index $b$ of $\rho_b$ and $m_b$).
The Ricci tensor for the diagonal metrics
$$ g_{ik}=e_i e^{2F_i} \delta_{ik}, \; e_i=(1,-1,-1,-1),$$
can be easily calculated using formulae
$$ R_{ii}=\sum\limits_{l\ne i}\Bigl[F_{i,i}F_{l,i}-F_{l,i}^2-F_{l,i,i}+e_i
e_le^{2(F_i-F_l)}(F_{l,l}F_{i,l}-F_{i,l}^2-F_{i,l,l}-F_{i,l}\sum\limits_{m\ne
i,l}F_{m,l})\Bigr],$$
$$ R_{ik}=\sum\limits_{l\ne
i,k}\Bigl[F_{l,k}F_{k,i}+F_{i,k}F_{l,i}-F_{l,i}F_{l,k}-F_{l,i,k}\Bigr], \;i\ne k.$$
Finally the Einstein equations yield:
$$ -8\pi T^R_R=e^{-2\lambda}\left( {r'^2\over r^2}+2{r'\over r}\phi'\right)
-{1\over
r}e^{-2\phi}\left(2(\ddot r-\dot r\dot\phi)+{\dot r^2\over r}\right)-{1\over r^2}, \eqno(5.1)$$
$$ -8\pi T^\theta_\theta=e^{-2\lambda}\left(\phi''+\phi'^2+{r''\over r}+{r'\over
r}(\phi'-\lambda')\right)-e^{-2\phi}\left(\ddot\lambda+\dot\lambda^2+{\ddot r\over r}+{\dot
r\over r}(\dot\lambda-\dot\phi)\right), \eqno(5.2)$$
$$ 8\pi T^t_t=-e^{-2\lambda}
\left(2{r''\over r}+{r'^2\over r^2}-2{r'\lambda'\over r}\right)+
e^{-2\phi}\left(\dot
{r^2\over r^2}+2{\dot\lambda\dot r \over r}\right)+{1\over r^2}, \eqno(5.3)$$
$$ 8\pi T^R_t=e^{-2\lambda}{2\over r}\left(\dot r'-\phi'\dot r - \dot\lambda
r'\right).\eqno(5.4)$$
The hydrodynamical momentum-energy tensor
$$ T_{\mu\nu}=(\eps+p)u_\mu u_\nu-p g_{\mu\nu} $$
has very simple mixed components in the comoving frame:
$$ T^t_t=\eps, \quad T^i_i=-p, i=1,2,3, \quad T^i_j=0, i\ne j $$
which should be inserted in the LHS of eqns. (5)

Let us also write down the conservation laws (which follow from Einstein
eqns(5) because the 4-divergence of Einstein tensor equals 0)
$$ T^\mu_{t;\mu}=\dot\eps+(\eps+p)(\dot\lambda+2{\dot r\over r})=0,
\eqno(6.1)$$
$$ T^\mu_{R;\mu}=-p'-\phi'(\eps+p)=0.\eqno(6.2)$$
We used here the following useful formula for divergence of a tensor:
$$ T^\mu_{\nu;\mu}={1\over\sqrt{-g}}\left(\sqrt{-g}T^\mu_\nu
\right)_{,\mu}-{1\over 2}g_{\mu\sigma,\nu}T^{\mu\sigma}.$$
From the comparison of eqns(6.1) and (4) we see that
$$ {\dot\eps\over\eps+p}={\dot\rho\over \rho}, \eqno(7) $$
which is nothing more than the adiabaticity relation $dE+pdV=0$.
Using this relation we also get the formula for the internal energy $\eps_{in}$ defined
as ${\eps=\rho(1+\eps_{in})}$:
$$ \dot \eps_{in}={\d \over\d  t}({\eps\over\rho})=-p{\d \over\d
t}({1\over\rho}). \eqno(8)$$

Now let's define the following quantities: $$ \tilde m \equiv \int 4\pi r^2\eps
{\d  r\over\d  R} dR,\eqno(9)$$ $$ U\equiv e^{-\phi} \dot
r,\eqno(10)$$ $$ \Gamma \equiv e^{-\lambda} r'.\eqno(11)$$ From this
definition we simply find using the relation (6.1) for $\dot\eps$
with excluded $\dot\lambda$ via (5.4) and
expressing $\phi'$ through (6.2): $$ \dot{\tilde m}=-4\pi r^2 p \dot r \eqno(12)$$
Equation (5.3) with $T^t_t=\eps$ provides further relation between $\tilde m(R,t)$
and $\Gamma,U$. Really, $$8\pi\eps r^2r'=(r-e^{-2\lambda}rr'^2+e^{-2\phi}r\dot
r^2)',$$ and hence
$$ 2\tilde m(R,t)=\int\limits_0^R8\pi r^2r'\eps dR=r(1+U^2-\Gamma^2). \eqno(13)$$
We do not have to forget the relation (3) which reads
$$ \Gamma/\rho={\d \over\d  m_b}(\frac{4\pi}{3}r^3), \eqno(3')$$
or equivalently
$$ \tau\equiv{1\over \rho}=4\pi r^2e^\lambda.\eqno(3'')$$

Now we are able to formulate equations for evolution of $\rho, U,
\eps_{in}$. Using (7) and (6.1),
$$ \dot\tau\equiv{\d \over \d  t}{1\over\rho}={1\over\rho}\left({2\dot r\over
r}+{\dot r'-\dot r\phi'\over r'}\right)=e^\lambda 4\pi r^2\left({2\dot r\over
r}+{\dot r'-\dot r\phi'\over r'}\right)={e^\lambda\over r'}(4\pi r^2\dot r
e^{-\phi})'e^{\phi}={1\over\Gamma}(4\pi r^2 U)'e^\phi,$$
i.e. finally
$$ e^{-\phi}{\d \over \d  t}{1\over\rho}={1\over\Gamma}(4\pi r^2
U)'.
\eqno(14)$$
Going further,
$$ e^{-\phi} \dot U=e^{-2\phi}(\ddot r-\dot
r\dot\phi)=e^{-2\lambda}({r'^2\over 2r}+2r'\phi')-4\pi rp-{1\over
2r}-e^{-2\phi}{\dot r^2\over 2r}, $$
where we have used (5.1) for  expressing
${(\ddot r -\dot{r}\dot{\phi})e^{-2\phi}}$,
and finally
$$ e^{-\phi} \dot U=e^{-2\lambda}2r'\phi'-4\pi rp-{\tilde m\over r^2}=
-{\Gamma\over h}4\pi r^2 p' -{\tilde m\over r^2}-4\pi rp, \eqno(15)$$
where we used the relation (13) for $\tilde m/r$ and introduced the specific
enthalpy
$$ h=(\eps+p)/\rho. \eqno(16)$$
The set of equations (14,15,8) can be used as hydrodynamical equations
without ultrarelativistic particles' fluxes (i.e., additional terms in the
energy-momentum tensor). Let me mention that Einstein equations (5.2,5.3)
were not used {\it directly} in the derivation, but instead two conservation
laws (6.1,6.2) were used which are themselves consequences of the whole set (5).

Now let's add the  momentum-energy tensor of ultrarelativistic particles,
let us call them ``neutrino'' (but these may also be photons)
$$ T^{(\nu)}_{\mu\nu}=(\rho E_\nu +p_\nu)u_\mu u_\nu +
(F^{(\nu)}_\mu u_\nu+F^{(\nu)}_\nu u_\mu) -p_\nu g_{\mu\nu}. \eqno(17) $$
$p_\nu$ is the neutrino pressure (not a 4-vector).

We may calculate the divergence of a slightly more general tensor
$$T_{\mu\nu}=A u_\mu u_\nu +
(B_\mu u_\nu+B_\nu u_\mu) +C g_{\mu\nu}, \eqno(18) $$
which yields
$$ T^\nu_{t;\nu}=\dot A+\dot C+({2\dot r\over r}+\dot\lambda)A+
\left[\left(\dot B_0 +B_0({2\dot r\over r}+\dot\lambda)\right)(1+e^{-\phi})
-B_0\dot\phi+B_R'+B_R({2r'\over r}+\phi'+\lambda')\right], \eqno(19.1)$$
$$T^\nu_{R;\nu}=C'-\phi'(A+2B_0)+e^{-\phi}\left(\dot B_R+B_R({2\dot r\over
r}+\dot\lambda)\right).\eqno(19.2)$$
Here $B_R=g_{RR}B^R=-\gamma_{RR}B^R=-e^{2\lambda}B^R$.
In our case $A=\eps+p+\rho E_\nu+p_\nu, B_\mu=F^{(\nu)}_\mu, C=-(p+p_\nu).$
We also introduce notations $Q,q$ according to:
$$ T^{(\nu)\alpha}_{\beta;\alpha}=(e^\phi Q,-e^\lambda q, 0 , 0), \quad
T^{(\nu)\alpha\beta}_{\; ;\alpha}=(e^{-\phi} Q, e^{-\lambda} q, 0 , 0),
\eqno(20)$$
so that $Q,q$ may be expressed through $\eps_\nu\equiv \rho E_\nu,\quad p_\nu,\quad F_\nu\equiv \gamma_{RR}F^{(\nu)R}$
(4-vector of flux must have $F_0=0$ in the local rest frame):
$$ e^\phi Q=\dot\eps_\nu+(\eps_\nu+p_\nu)(\dot\lambda+{2\dot r\over
r})-F'_\nu-F_\nu(\phi'+\lambda'+{ 2r'\over r}),$$
$$ e^\lambda q=p'_\nu+\phi'(\eps_\nu+p_\nu)-e^{-\phi}\left(\dot
F_\nu+F_\nu(\dot\lambda+{2\dot r\over r})\right);\quad  \dot\lambda+{2\dot r\over
r}=-{\dot\rho\over\rho}.$$

The components of momentum-energy tensor are now
$$ T^t_t=\eps+\rho E_\nu; 
\quad T^R_t=e^\phi F^{(\nu)R}=e^{\phi-2\lambda}F_\nu;\quad
T^R_R=T^\theta_\theta=-(p+p_\nu).$$
From (5.4) we obtain now
$$ \dot \lambda={\dot r'-\phi'\dot r-4\pi r e^{\phi}F_\nu\over r'},
\eqno(5.4^\nu)$$
and (6) are replaced by
$$ \dot\eps=-(\eps+p)(\dot\lambda+{2\dot r\over r})-e^\phi
Q,\eqno(6.1^\nu)$$
$$ \phi'=-{p'+e^\lambda q\over \eps+p}=-{\tau\over h}p'-{\tau^2\over 4\pi
r^2}{q\over h}. \eqno(6.2^\nu)$$
Some minor changes in other relations:
$$ h\dot\rho=\dot\eps+e^\phi Q \eqno(7^\nu)$$
$$ e^{-\phi}\dot\eps_{in}=-p e^{-\phi}\dot\tau-\tau Q \eqno(8^\nu)$$
$$ \tilde m(R,t)=\int 4\pi r^2 T^t_t r' dR \eqno(9^\nu)$$
Equations (10,11,13) are not modified, but (14,15) acquire additional
neutrino couplings in the RHS:
$$ e^{-\phi}\dot\tau={1\over\Gamma}(4\pi r^2
U)'-\tau{4\pi r\over \Gamma }F_\nu,
\eqno(14^\nu)$$
$$ e^{-\phi} \dot U=
-{\Gamma\over h}4\pi r^2 (p'+{\tau\over 4\pi r^2}q) -{\tilde m\over r^2}-
4\pi r(p+p_\nu). \eqno(15^\nu)$$

\section{Lagrangean special relativistic hydrodynamics}
After having formulated hydrodynamical equations in the comoving RF, we need to relate
some quantities with observable values in some fixed RF. We'll do this in the limit of
very small gravitational mass, when a pseudo-euclidean (Minkowskian) RF exists:
$$ ds^2= d\tilde t^2 - dr^2 - r^2 d\Omega^2. \eqno(21)$$
It is obvious that $r$ here is the same as in the comoving RF (1), because the length
of a circle ($2\pi r$) is the same in both RF's (Lorentz transformation does not change
transverse sizes, and matter moves only radially). For the same reason
$d\Omega^2$ is the same in both RF's, because it is the area on the unit
sphere which is not transformed being transverse to the motion.

We need to relate times $t$ and $\tilde t$, and this must be some function $\tilde t(t,m)$. Derivatives
of this function can be easily determined. As for a moving particle (which has $m=$ const), its
worldline interval equals $ds=e^\phi dt$, and on the other hand,
$$ds=
\sqrt{d\tilde t^2-\left({\d  r\over \d  \tilde t}\right)^2_m d\tilde t^2}=d\tilde t \sqrt{1-v^2}={1\over\gamma}d\tilde t,$$
where we've defined $v=\left({\d  r\over\d  \tilde t}\right)_m,$ which is simply radial velocity
of a particle $(m)$ in the fixed RF, and usual Lorentz factor $\gamma={1\over\sqrt{1-v^2}}$.
Hence,
$$ \left({\d\tilde t\over\d  t}\right)_m=\gamma e^{\phi(m,t)}. \eqno(22)$$

Now determine $ \left({\d  \tilde t\over\d  t}\right)_r.$
 For a stationary point ($r=$ const)
we have ${ds^2=d\tilde t^2=e^{2\phi}dt^2-e^{2\lambda}dm^2,}\quad$
 $ {dm=-{\d  r/\d  t\over\d  r/\d  m}}dt$
 (derivative along the curve $r(m,t)=$ const).
Therefore,
$$\left({\d  \tilde t\over\d  t}\right)_r=
\sqrt{e^{2\phi}-e^{2\lambda}\left({\d  r/\d  t\over\d  r/\d  m}\right)^2}.
\eqno(23)$$
And with this formula it is straightforward to find $\left({\d  \tilde t\over\d  m}\right)_t$:
as $\left({\d  \tilde t\over\d  t}\right)_r=
\left({\d  \tilde t\over\d  t}\right)_m+\left({\d  \tilde t\over\d  m}\right)_t
\left({\d  m\over\d  t}\right)_r,$
$$\left({\d  \tilde t\over\d  m}\right)_t=
{\d  r/\d  m\over\d  r/\d  t}
\left(\left({\d  \tilde t\over\d  t}\right)_m-
\left({\d  \tilde t\over\d  t}\right)_r\right)
={e^{\lambda}\Gamma\over e^{\phi}U}
\left(\gamma e^\phi-\sqrt{e^{2\phi}-\left({e^\phi U\over\Gamma}\right)^2}\right).\eqno(24)$$
We recall that $U=\gamma v$
(this follows from the equality $U=\left({\d  r\over \d  s}\right)_m,
\quad ds={1\over\gamma}d\tilde t$ for a moving particle).
Moreover, in the case of weak gravitational fields which we assumed,
$$ \Gamma=\sqrt{1+U^2-{2\tilde m\over r}}\approx\gamma-{\tilde m\over\gamma r}\approx\gamma.\eqno(25)$$
The difference between $\Gamma$ and $\gamma$, that is, ${\tilde m\over \gamma r}$, is always small
in the ultrarelativistic case: it goes like $\sim \rho_{tot} r^2/\gamma$ for small $r$, equals
${G M_{tot}\over\gamma r}$ outside of the most mass of matter, and never exceeds about ${G M_{tot}\over\gamma R}\sim{R_{grav}\over\gamma R}$.
So, derivatives (23,24) take the form:
$$ \left({\d  \tilde t\over\d  t}\right)_{r}={1\over\gamma}e^\phi,\quad\left({\d  \tilde t\over\d  m}\right)_t=e^\lambda U.\eqno(26)$$
And when one wishes to express $(t,m)$-derivatives through $(\tilde t,m)$-derivatives, the next formulae are to be applied:
$$\left({\d \over \d  t}\right)_m=
\gamma e^\phi \left({\d  \over\d  \tilde t}\right)_m,\eqno(27.1)$$
$$\left({\d  \over\d  m}\right)_t=
\left({\d  \over\d  m}\right)_{\tilde t}
+e^\lambda U \left({\d  \over\d  \tilde t}\right)_m.\eqno(27.2)$$
We also write down for reference the formula relating distances in the two
RFs:
$$ dr|_{\tilde t}={1\over\gamma^2}dr|_t.$$
(And remember that physical length between two mass points in the comoving
RF is $dL=e^{\lambda}dm={1\over\gamma}dr|_t;$ it is now clear that the distance between the same points
measured at the moment $\tilde t$ in the euclidean RF is just $dL/\gamma$.)
One is able to see now, that from the $m$-derivatives of pressure come also $\tilde t$-derivatives,
and this circumstance complicates the set of equations. For example, rewrite
the equation (14) for density evolution (for simplicity without photon/neutrino flux)
in $(\tilde t,m)$-terms:
$$ \gamma {\d \tau\over\d  \tilde t}={4\pi\over\gamma}\left[{\d \over\d
m}\left(r^2U\right)_{\tilde t}+e^\lambda U{\d \over\d
\tilde t}\left(r^2U\right)\right]={4\pi\over\gamma}\left[{\d \over\d
m}\left(r^2U\right)_{\tilde t}+e^\lambda U\left(2r(\gamma-1)+r^2{\d
U\over\d  \tilde t}\right)\right].\eqno(28)$$
We can express here ${\d
U\over\d  \tilde t}$ using the second equation of our set (15), which reads
in new variables
$$\gamma {\d  U\over\d  \tilde t}=-4\pi r\left[  p + {\gamma\over
h}r\left\{\left({\d  p\over\d  m}\right)_{\tilde t}+e^\lambda U{\d
p\over\d  \tilde t}\right\}\right].\eqno(29)$$
We see that some additional  $\tilde t$-derivatives appear on the RHS of evolutionary
equations for (inverse) density, but fortunately this can  be avoided writing equations for
${1\over\gamma\rho}$ instead of ${1\over\rho}$. Really, since
$d\gamma=v d(\gamma v),$ we find
$$ {\d  \over\d  \tilde t}{1\over
\gamma\rho}-{1\over\gamma^2}{\d \over\d  m}(4\pi r^2
v)={4\pi r^2v\over\gamma^3}{\d \gamma\over\d  m}+{2v^3\over
\rho\gamma r},$$
i.e., introducing $V_m\equiv{1\over\rho\gamma}$ and denoting $\d/\d \tilde t$ and
$\d/\d m$ by dot and prime, respectively,
$$\dot V_m-{4\pi\over\gamma^2}(r^2v)'={4\pi r^2v\gamma'\over\gamma^3}+{2v^3\over
\rho\gamma r}.$$
Substituting here $\gamma'=\gamma^3 v v'$ and bearing in mind that
$$\left({\d r\over\d m}\right)_{\tilde t}={1\over\gamma}e^\lambda={1\over
4\pi r^2\gamma\rho}$$
(but compare this with eqn. (11) --- this is {\it different} derivative because it is taken at constant $\tilde t$)
we finally see
$$\dot V_m-4\pi(r^2v)'=0.\eqno(30)$$
 The same trick may be applied to evade $\d/\d \tilde t-$derivatives in the RHS of
eqn.
for momentum density $U$ if one uses $S_m\equiv hU$ instead of $U$:
$$\dot S_m+4\pi r^2p'=-{4\pi rph\over\gamma}.\eqno(31)$$
For reference let me write down also the very equation for $U$:
$$\dot U=-{4\pi r^2\over h}(p'+{U\over 4\pi r^2\rho}\dot p)-{4\pi
rp\over\gamma}.\eqno(29')$$
Using $E_m\equiv h\gamma-{1\over\gamma}{p\over\rho}$
as the third variable (instead of $\eps_{in}$) we also do get a total cancellation
of $\tilde t-$derivatives (one should also express $\dot p$ through $\dot U$ and $\dot\gamma\equiv v\dot U$):
$$\dot E_m+4\pi (r^2\gamma vp)'=-4\pi {rvhp\over\gamma}.\eqno(32)$$

Let me also
mention that if we had not put $G=1$, the only additional multiple in the set (30-32) would be
$G$ in the RHS of (31) and (32). This means that these additional (with
respect to basic eqns. [2]) terms are of the order $O({G M\over\gamma R})$
and therefore must be neglected since we have already omitted such terms,
for example, in (25). Moreover, to take into account these corrections, coming from
General relativity,  one should change the very assumption of existence of
the metrics (21).

 Finally, let's introduce one "new" independent variable, following the notations of
the work [2]:
$$ d\bar m=\gamma\rho\; 4\pi r^2 dr|_{\tilde t}\equiv{\rho\over\gamma}\; 4\pi r^2 dr|_t.\eqno(33)$$
Clearly, $d\bar m|_{\tilde t}=dm|_t$, so that $\bar m$ is  a baryon rest-mass Lagrangean
coordinate. Actually, $\bar m(\tilde t,r)$ defined as
$$ \bar m(\tilde t,r)=\int\limits_{r_0}^r{\gamma\rho\; 4\pi r^2\; dr|_{\tilde t}}$$
is the same baryon rest mass like in the comoving RF but expressed as a function of $(\tilde t,r)$:
$$ \bar m(\tilde t,r)\equiv m(t,r). $$
This implies that our notations are identical to those in [2].
From this definition of $m$ it is easy to derive an equation for $\dot V_m$:
$$ \left({\d m\over \d r}\right)_{\tilde t}={4\pi r^2\over V_m},$$
$$ \dot V_m=4\pi {\d\over\d\tilde t}(r^2 {\d r\over\d m})=4\pi(2r\dot r {\d r\over\d m} + r^2
{\d^2 r\over\d\tilde t\d m})=4\pi(2rv{\d r\over\d m} + r^2{\d v\over\d m})={\d \over\d m}(4\pi r^2 v).$$

\section{Conclusions}

Let us summarize our results. If one is to consider General Relativistic
effects, the exact set of hydrodynamical equations (14,15,8) can be used.
In the weak field limit, one can apply the easier (and more suitable for comparison with observational
data) set of Special Relativistic equations (30-32) which neglect $O(GM/R)$ effects
but are formulated in the fixed pseudoeuclidean spacetime, call this ``lab
frame''. But if an exact set is used, it is desirable nevertheless to
present results in $(r,\tilde t)$-description, where $\tilde t$ is the
universal time in lab-(observer-)frame. To establish this, the functions
$\tilde t(t,m)$ and $r(t,m)$ should be found, e.g., via integrating together
with a set of hydro equations the following 4 equations:
$$ {\d\tilde t\over\d t}=e^{\phi}\gamma,\quad{\d\tilde t\over \d m}=e^\lambda U;$$
$$ {\d r\over\d t}=e^\phi U,\quad{\d r\over \d m}=e^\lambda \gamma.$$
This Jacobian matrix  ${\d (\tilde t, r)\over \d (t,m)} $ has an evident remarkable symmetry.

So, if one knows a GR-solution, for instance, a path of a shock wave,
$m_{SW}(t)$, then in the lab-frame the solution is a  curve
$r=r(t,m_{SW}(t)),\quad \tilde t=\tilde t(t,m_{SW}(t))$
parameterized by $t$.

\section*{Acknowledgement}
I would like to thank Serguei Blinnikov for stimulating discussions and
corrections to the text.

This work is partially supported by {\bf RFBR Grant 99-02-16205}.

\vskip 2cm
\section*{ References}
\par\indent

[1] {\it S. Yamada.} An implicit Lagrangian code for spherically symmetric GRH... // ApJ {\bf 475} p.720

[2] {\it F. Daigne, R. Mochkovitch.} GRBs from internal shocks in a relativistic wind: an hydrodynamical study
// published in A\&A, see {\tt astro-ph/0005193}

[3] {\it L. Wen  et al.} ApJ 486 p.919

[4] {\it R.D. Blandford, C.F. McKee.} Fluid dynamics of relativistic blast waves
//Phys. of Fluids, Vol.19 no. 8 p.1130 (1976)

[5] { \it M. Liebend\"orfer, A. Mezzacappa, F.-K. Thielemann } Conservative
General Relativistic Radiation Hydrodynamics in Spherical Symmetry and
Comoving Coordinates // {\tt astro-ph/0012201}

\end{document}